# DLBayesian: An Alternative Bayesian Reconstruction of Limited-view CT by Optimizing Deep Learning Parameters

Changyu Chen, Li Zhang, Yuxiang Xing, and Zhiqiang Chen

*Abstract*—Limited-view computed tomography (CT) presents significant potential for reducing radiation exposure and expediting the scanning process. While deep learning (DL) methods have exhibited promising results in mitigating streaking artifacts caused by a reduced number of projection views, their generalization remains a challenge. In this work, we proposed a DL-driven alternative Bayesian reconstruction method (DLBayesian) that efficiently integrates data-driven priors and data consistency constraints. DLBayesian comprises three stages: group-level embedding, significance evaluation, and individual-level consistency adaptation. Firstly, DL network parameters are optimized to learn how to eliminate the general limited-view artifacts on a large-scale paired dataset. Then, we introduced a significance score to quantitatively evaluate the contribution of parameters in DL models as a guide for the subsequent individual-level adaptation. Finally, in the Bayesian adaptation stage, an alternative Bayesian reconstruction further optimizes the DL network parameters precisely according to the projection data of the target case. We validated DLBayesian with sparse-view (90 views) projections from a circular trajectory CT and a special data missing case from a multi-segment linear trajectory CT. The results underscore DLBayesian's superior generalization capabilities across variations in patients, anatomic structures, and data distribution, as well as excelling in contextual structure recovery compared to networks solely trained via supervised loss. Real experiments on a dead rat demonstrate its capability in practical CT scans.

## I. INTRODUCTION

COMPUTED tomography (CT) is a non-invasive imaging technique that finds widespread usage across clinical, security, and industrial applications[1-3]. To achieve high-quality reconstruction, projection data should be densely sampled over a sufficient angular range[4]. However, current advancements in radiation dose reduction, dynamic imaging, and non-standard scan trajectory have spurred the relevance of limited-view CT (LVCT) scans. By LVCT，we refer to scans only offering a fraction of projection views which mainly include two pathways: **(1) Sparse-view CT** reduces sampling density within sufficient angular coverage as a common practice in radiation dose reduction[5]; **(2) Limited-angle CT** restricts angular coverage due to time constraint in dynamic imaging or spatial limitation for scanning large objects[6]. Occasionally, the two pathways may intertwine due to unique configurations, such as stationary CTs, resulting in a complex mixed LVCT problem[7, 8].

LVCT reconstruction is a challenging ill-posed problem. Conventional analytical techniques often yield degraded image quality with streaking artifacts and structural distortions. Leveraging the Bayesian theory, model-based iterative reconstruction (MBIR) integrates statistical properties, image priors, and imaging physics to enhance reconstructions. These methods typically formulate a joint optimization problem with the data fidelity term and an image sparsity prior, such as total variation[9], dictionary learning[10], and low rank[11]. However, due to the limited precision and capability of manually crafted priors, compromised image quality and new artifacts may appear in practical applications.

In these years, deep learning (DL) methods have demonstrated strong capability in resolving numerous ill-posed problems[12-14]. These methods train neural networks (NN) on a large-scale dataset to extract shared features and pose implicit priors for the reconstructions. In the context of LVCT reconstruction, they predominantly fall into three categories: **(1) Single-domain methods** employ NNs for sinogram inpainting to obtain fully-sampled projections[15, 16], or for image post-processing to eliminate the artifacts in the degraded images [17, 18]. While these methods offer computational efficiency, they fail to integrate the imaging models and consistency with the projection data. **(2) Cross-domain methods** conduct a more comprehensive optimization between the projection domain and image domain with the reconstruction process embedded as a domain transformer[19, 20]. Our group proposed Sam's Net which employs a self-augmented mechanism to broaden the distribution of dataset based on the imaging model[21]. **(3) Unrolled and multi-stage methods** iteratively leverage the strength of the data consistency and data-driven priors[22-28]. Recently, there are also other techniques that got a lot of attention in this field. Implicit neural representation employs multi-layer perceptron to learn a mapping function from coordinates to intensity based on individual projection data in a

This work was supported in part by the National Natural Science Foundation of China (Grant No. 12327809). Changyu Chen, Yuxiang Xing, Li Zhang and Zhiqiang Chen are with the Department of Engineering Physics, Tsinghua University, Beijing, 100084, China, and are also with the Key Laboratory of Particle & Radiation Imaging (Tsinghua University), Ministry of Education, Beijing, 100084, China (email: ccy, zli, xingyx, czq@mail.tsinghua.edu.cn). **Corresponding authors: Yuxiang Xing; Zhiqiang Chen**.

self-supervised manner which is similar to iterative reconstruction [29, 30]. Diffusion models usually work on a large-scale dataset and incorporate a forward process to perturb data to pure noise and a reverse process to convert noise back to the final reconstruction[31-34].

DL methods basically work under the assumption of a similar distribution between the training data (source domain) and real-world data (target domain)[35]. Nevertheless, the precise simulation of the object anatomy, scan protocols, imaging geometries, and noise distributions in practical CT scans is challenging, especially for new imaging systems and clinical applications. This discrepancy leads to domain shift and generalization problems[36]. In the literature, domain adaptation methods are widely investigated[37-40]. For CT reconstruction, it is worth noting that the acquired projection data reflect valuable information for individual cases in the target domain. This potentially provides a self-supervised signal for knowledge transfer. Conventional MBIR methods could incorporate the outputs from DL models either as an enhanced initial point or a prior image to efficiently constrain the problem[41, 42]. As the network parameters are frozen in the inference phase, its adaptability to large domain shifts may be limited. Consequently, it is crucial to identify an efficient pathway to harness the strength of both the individual-related projection data and the group-defined prior knowledge learned by the NNs. In this work, we present a novel approach, termed DLBayesian, which unites **D**eep-**L**earning priors with data consistency constraints via an alternative **Bayesian** reconstruction framework. In practice, it works in a three-stage manner: a prior model training stage where the DL network learns to eliminate artifacts on a large-scale paired dataset, a significance evaluation process that quantifies the contribution of network parameters using a proposed significance score metric, and a consistency adaptation stage where the significant part of DL parameters is finetuned to accommodate each individual case based on the Bayesian theory.

Our main contributions are summarized as follows:

(1) We established a novel alternative Bayesian reconstruction framework that combines the strength of DL and CT imaging models to achieve high-quality reconstruction for personalized X-ray CT data.

(2) We introduced a significance score to quantitatively evaluate the contribution of parameters in DL models and revealed its correlation with generalization. The score serves as an important criterion to effectively and robustly guide the proposed DLBayesian process.

(3) The proposed DLBayesian method is extensively validated on simulated and real experiments across multiple scan trajectories and anatomic structures which demonstrates great flexibility in various scenarios and capability of handling practical CT scans.

The rest of this paper is organized as follows. The problem formulation of LVCT and DLBayesian framework is introduced in Section II. In Section III, the experimental results are demonstrated. Finally, the discussion and conclusion are given in Sections IV and V.

## II. METHODS

### A. Formulation of Limited-view CT

We formulate a full-view projection $\mathbf{p}_{FV} \in \Re^{N_\theta N_D}$ as

$$\mathbf{p}_{FV} = \mathbf{H}_{FV}\boldsymbol{\mu} + \mathbf{n}_{FV} \quad (1)$$

with $N_\theta$ and $N_D$ being the number of projection views and detector elements respectively. $\mathbf{H}_{FV} \in \Re^{N_\theta N_D \times N_W N_H}$ denotes the system matrix, $\boldsymbol{\mu} \in \Re^{N_W N_H}$ the imaged object and $\mathbf{n}_{FV} \in \Re^{N_\theta N_D}$ the statistical noise. Given that only partial projection views $N'_\theta$ were acquired, one obtains LVCT projection data $\mathbf{p}_{LV} \in \Re^{N'_\theta \times N_D}$. Based on the distribution of projection views, LVCT problems can be divided into two main types: (1) sparse-view CT with reduced sampling density (see **Fig. 1**(a)); (2) limited-angle CT with restricted angular coverage (see **Fig. 1**(b)).

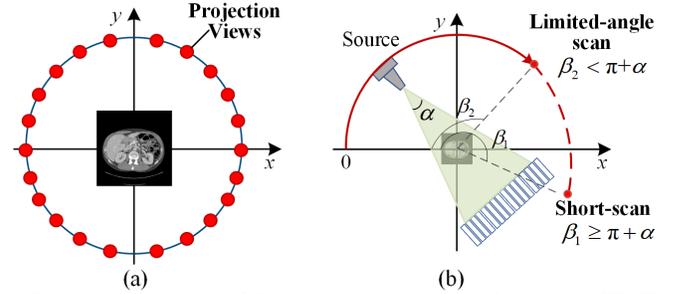

**Fig. 1** Illustrations of limited-view scans in a fan beam CT: The imaging geometries of sparse view CT (a) and limited-angle CT (b).

### B. Existing Bayesian Reconstruction

A Bayesian reconstruction method reconstructs images with the maximum posterior from LVCT projection $\mathbf{p}_{LV}$ as

$$\hat{\boldsymbol{\mu}} = \arg\max_{\boldsymbol{\mu} \geq 0} \text{Prob}(\boldsymbol{\mu} | \mathbf{p}_{LV}) = \arg\max_{\boldsymbol{\mu} \geq 0} \frac{\text{Prob}(\mathbf{p}_{LV}|\boldsymbol{\mu})\text{Prob}(\boldsymbol{\mu})}{\text{Prob}(\mathbf{p}_{LV})} \quad (2)$$

where $\text{Prob}(\cdot)$ is the probability distribution function. After negative log transformation, it can be simplified to

$$\hat{\boldsymbol{\mu}} = \arg\min_{\boldsymbol{\mu} \geq 0} \psi_{Fid}(\boldsymbol{\mu}, \mathbf{p}_{LV}) + \lambda \psi_{Prior}(\boldsymbol{\mu}) \quad (3)$$

Here, $\psi_{Fid}(\boldsymbol{\mu}, \mathbf{p}_{LV}) = -\ln \text{Prob}(\mathbf{p}_{LV}|\boldsymbol{\mu})$ is the data fidelity that measures the consistency between a projection and the corresponding reconstruction. It often appears in a weighted least square format $\psi_{Fid}(\boldsymbol{\mu}, \mathbf{p}_{LV}) = \|\mathbf{p}_{LV} - \mathbf{H}_{LV}\boldsymbol{\mu}\|_{\Sigma}^2$ with $\Sigma$ being the noise covariance and $\mathbf{H}_{LV}$ the system matrix for LVCT scan. $\psi_{Prior}(\boldsymbol{\mu}) \propto -\ln \text{Prob}(\boldsymbol{\mu})$ is the regularization term with a hyper-parameter $\lambda$ leveraging its strength. As LVCT reconstruction is an underdetermined problem, there exists a null space satisfying $\psi_{Fid}(\boldsymbol{\mu}, \mathbf{p}_{LV}) = 0$. If we start from an initial estimation $\hat{\boldsymbol{\mu}}_{Ini}$ from analytical methods and purely rely on $\psi_{Fid}(\boldsymbol{\mu}, \mathbf{p}_{LV})$ ($\lambda = 0$), we may land on a point far from the ground truth $\boldsymbol{\mu}_{GT}$. With the prior knowledge of imaged objects ($\lambda > 0$), the solution could be closer to $\boldsymbol{\mu}_{GT}$. Manually designed *priori* is helpful but may not be efficient enough and compromise the resultant image quality.

### C. Design Rationale

DL has shown great potential in embedding and expressing implicit properties within extensive samples $\Omega$. We can divide $\Omega$ into a training set $\Omega_{Train}$, and a validation set $\Omega_{Val}$ to train a network $\varphi(\cdot)$ parameterized by $\vartheta$. In the training phase, $\vartheta$ is optimized to learn the mapping from $\hat{\boldsymbol{\mu}}_{Ini}$ to $\boldsymbol{\mu}_{GT}$ for all samples in $\Omega_{Train}$, and evaluated on $\Omega_{Val}$ to identify the model that best generalizes to unseen data. This model, known as the best validation (BV) model $\vartheta^{BV}$, uses the early stopping mechanism to mitigate overfitting. In the test phase, the BV model processes real data with frozen parameters representing the knowledge learned from $\Omega_{Train}$. If the real data distribution is similar to $\Omega_{Train}$, the BV model can achieve high-quality reconstruction by $\hat{\boldsymbol{\mu}}_{Net} = \varphi(\hat{\boldsymbol{\mu}}_{Ini}; \vartheta^{BV})$. However, in practical applications, variations in patients, anatomic structures, and sampling distribution could add to the difference between the training data and real data which leads to a suboptimal $\hat{\boldsymbol{\mu}}_{Net}$. One solution for this issue is to apply the Bayesian framework that leverages the individual-specific information under the Radon transform provided by the projection data as

$$\hat{\boldsymbol{\mu}} = \arg\min_{\boldsymbol{\mu} \geq 0} \|\mathbf{p}_{LV} - \mathbf{H}_{LV}\boldsymbol{\mu}\|_{\Sigma}^2 + \lambda \|\boldsymbol{\mu} - \hat{\boldsymbol{\mu}}_{Net}\|^2 \quad (4)$$

where $\hat{\boldsymbol{\mu}}_{Net}$ is either an initial point or a prior image. While this pipeline demonstrates effectiveness, its performance largely depends on the prior information derived from $\hat{\boldsymbol{\mu}}_{Net}$ and the setting of hyper-parameters in the Bayesian reconstruction.

Upon reviewing the working mechanism of DL networks, the prior information is encoded within the NN function $\varphi(\cdot)$ and its parameters $\vartheta$ while $\hat{\boldsymbol{\mu}}_{Net}$ is merely a specific data point of $\varphi(\cdot)$ tailored to a particular sample. Hence, one natural idea is to fully integrate the DL model into the Bayesian reconstruction framework and directly optimize in the DL parameter space to obtain a modified NN function. Specifically, we employ a "group-to-individual" strategy: the first step (termed "prior model training") learns a NN function $\varphi(\cdot)$ with its parameters $\vartheta^{(0)}$ to extract group-level prior information on $\Omega_{Train}$, and the second step (termed "Bayesian adaptation") derives a new function parameterized by $\vartheta^*$ through precisely adapting the prior knowledge from $\vartheta^{(0)}$ to suit each specific test sample:

$$\hat{\boldsymbol{\mu}} = \varphi(\hat{\boldsymbol{\mu}}_{Ini}; \vartheta^*)$$

$$\text{s.t.} \quad \vartheta^* = \arg\min_{\vartheta} \|\mathbf{p}_{LV} - \mathbf{H}_{LV}\varphi(\hat{\boldsymbol{\mu}}_{Ini}; \vartheta)\|_{\Sigma}^2 + \lambda \psi_{Prior}(\varphi(\hat{\boldsymbol{\mu}}_{Ini}; \vartheta)) \quad (5)$$

In this way, the group-level prior information and individual-level projection data are unified under an alternative Bayesian reconstruction framework. Diverging from the existing image-space optimization methods, optimization in the DL parameter space exhibits two distinct advantages: firstly, it can directly inherit useful information from the training process by utilizing the same DL model architecture and suitable initialized weights; secondly, the output from deep neural networks tends to be smooth which intrinsically impose priors on ill-posed problems [43, 44]. To establish a robust adaptation procedure, two key aspects need to be addressed: firstly, it is essential to ensure that NNs comprehensively learn from $\Omega_{Train}$ to retrieve valuable prior knowledge; secondly, the group-level knowledge encoded in $\vartheta^{(0)}$ should be efficiently adapted to each individual case.

The first aspect pertains to the initialization for Bayesian adaptation. A BV model is a common choice to balance the trade-off between fitting $\Omega_{Train}$ and generalizing to unseen data. However, since the selection of BV model is based on the performance on $\Omega_{Val}$ during training, its effectiveness relies on the composition of $\Omega_{Val}$. Moreover, the early stopping mechanism often results in a BV model that is suboptimal for $\Omega_{Train}$. Our proposed "group-to-individual" strategy decouples the learning and generalization processes which mitigates concerns about overfitting. At the group-level learning stage, the primary goal is to achieve optimal performance under the selected network architecture and derive optimal weights for $\Omega_{Train}$. Subsequently, the Bayesian adaptation procedure focuses on the distinct features of the new data to enhance adaptation performance. Therefore, we propose to train the NN function to convergence and use the resulting optimal weights on $\Omega_{Train}$ as the initialization point for the alternative Bayesian reconstruction.

The second aspect emphasizes the efficiency and stability of the Bayesian adaptation process. This relies on a thorough analysis of the components of the learned NN function $\varphi(\cdot)$ as addressed in the following subsection.

### D. Evaluation of Network Parameter Significance

In the Bayesian adaptation stage, the most straightforward approach is to finetune all the NN parameters according to Eq.(5). However, the scale of the parameters directly influences the complexity of solving the optimization problem. NN models are often overparameterized to ensure effective fitting on large-scale datasets which results in redundant parameters when adapting to individual samples. In contrast, finetuning a subset of the model can enable more efficient and stable optimization but the selection of finetuned parameters is a challenge. Hence, we delve into the working mechanism of NNs when processing both training data and new data, assess the importance of model parameters to overall performance, and systematically reduce the parameter set to be finetuned. Here, we mainly focus on convolutional neural networks and treat each convolutional layer as a fundamental unit for strategy formulation. For a given convolution layer with weights $\mathbf{w}$, we evaluate its contribution to the whole loss function using a significance score $G_{\mathbf{w}}$ which is defined as the change in loss function $L(\vartheta, \Omega)$ due to a small perturbation in its weights.

$$G_{\mathbf{w}}(\vartheta, \Omega, L) \triangleq \left| L(\vartheta, \Omega) - L(\vartheta|_{\mathbf{w}'=\mathbf{w}+\Delta\mathbf{w}}, \Omega) \right|$$

$$\underset{\text{Taylor Expansion}}{\overset{\text{first-order}}{\approx}} \left| L(\vartheta, \Omega) - \left( L(\vartheta, \Omega) + \frac{\partial L(\vartheta, \Omega)}{\partial \mathbf{w}} \Delta\mathbf{w} \right) \right| \quad (6)$$

$$\underset{\Delta\mathbf{w}=c\mathbf{w}}{=\!=\!=} \left| c \frac{\partial L(\vartheta, \Omega)}{\partial \mathbf{w}} \mathbf{w} \right|$$

The quantity of $G_{\mathbf{w}}$ reflects the sensitivity of the NN function $\varphi(\cdot)$ to the changes in parameter weights. Assuming a linear perturbation, $G_{\mathbf{w}}$ can be approximately decomposed into three factors using a first-order Taylor expansion: a small constant $c$ that controls the magnitude of the perturbation, the gradient of $L(\vartheta,\Omega)$ to $\mathbf{w}$, and the weight $\mathbf{w}$. This indicates that the significance score of a convolution layer is mainly determined by the network architecture and weights $\vartheta$, the dataset $\Omega$, and the loss function $L(\cdot)$. The objective of Bayesian adaptation is to find appropriate $\vartheta^*$ for transferring knowledge from $\Omega_{\text{Train}}$ to a test sample. We found that the relative magnitude of significance scores for different convolution layers exhibit a consistent pattern across $\Omega_{\text{Train}}$ and unseen data $\Omega_{\text{Val}}$ (as illustrated in Section III.A). Therefore, $G_{\mathbf{w}}\left(\vartheta^{(0)},\Omega_{\text{Train}},L\right)$ is a suitable metric to guide the alternative Bayesian reconstruction.

### E. Framework Overview

By integrating data-driven DL networks with Bayesian adaptation, we propose our efficient **D**eep-**L**earning-prior based alternative **Bayesian** reconstruction (DLBayesian) which operates in a three-stage manner (as depicted in **Fig. 2**).

Firstly, in the prior model training stage, the DL network learns to extract the shared features across diverse contaminated and clean images, remove streaking artifacts through supervised learning, and achieve the best performance on $\Omega_{\text{Train}}$. In this way, NN fully captures the shared characteristics of relevant groups including artifact patterns associated with the scan configuration and anatomical structures.

$$\vartheta^{(0)} = \arg\min_{\vartheta} \sum_{j} \left\| \varphi(\hat{\boldsymbol{\mu}}_{\text{Ini}}^{(j)};\vartheta) - \boldsymbol{\mu}^{(j)} \right\|_{2} \quad (\{\hat{\boldsymbol{\mu}}_{\text{Ini}}^{(j)};\boldsymbol{\mu}^{(j)}\} \in \Omega_{\text{Train}}) \quad (7)$$

Secondly, in the significance evaluation stage, we consider a pretrained model $\vartheta^{(0)} = \{\mathbf{w}_m^{(0)}\}\big|_{m=1}^{M}$ with $M$ layers. The significance score $G_{\mathbf{w}_m}$ for the $m^{\text{th}}$ layer $\mathbf{w}_m^{(0)}$ can be computed by Eq. (6). Based on a predefined threshold $g$, all layers are divided into the significant part $\vartheta_{g+}$ and the less significant part $\vartheta_{g-} = \vartheta \setminus \vartheta_{g+}$ (referred to as the insignificant part).

$$\vartheta_{g+} = \left\{ \mathbf{w}_m \in \vartheta \mid G_{\mathbf{w}_m}\left(\vartheta^{(0)},\Omega_{\text{Train}},L\right) \geq g \right\} \quad (8)$$

Thirdly, in the Bayesian adaptation stage, only $\vartheta_{g+}$ are finetuned (using $\vartheta_{g+}^{(0)}$ as the initial point) and $\vartheta_{g-} = \vartheta_{g-}^{(0)}$ are fixed as a parametric prior to inherit the prior knowledge learned in stage 1. The optimization problem is formulated as

$$\hat{\boldsymbol{\mu}} = \varphi(\hat{\boldsymbol{\mu}}_{\text{Ini}};\vartheta^*)$$
$$\text{s.t.} \begin{cases} \vartheta^* = \arg\min_{\vartheta} \left\| \mathbf{p}_{\text{LV}} - \mathbf{H}_{\text{LV}}\varphi(\hat{\boldsymbol{\mu}}_{\text{Ini}};\vartheta) \right\|_{\Sigma}^{2} + \lambda \psi_{\text{Prior}}(\varphi(\hat{\boldsymbol{\mu}}_{\text{Ini}};\vartheta)) \\ \vartheta_{g-} = \vartheta_{g-}^{(0)} \end{cases} \quad (9)$$

Thus, the DL model is efficiently and robustly adapted to an individual unseen case, guided by prior knowledge and constrained by data fidelity.

## III. EXPERIMENTAL RESULTS

Limited-view projections were simulated under both a circular trajectory and a special multi-segment linear trajectory. We employed FBPConvNet as the backbone of $\varphi(\cdot)$ which consists of 23 convolutional layers (four of which are deconvolution layers for upsampling). In the prior model training stage, the network was trained on a simulated torso dataset derived from AAPM "Low Dose Grand Challenge" and TCIA "Low Dose CT Image and Projection Data" where the torso data from 38/11 patients were used for training/validation respectively. Test samples were from other patients covering the abdomen and head to thoroughly evaluate the performance of the proposed methods. The prior model was trained for 120 epochs in the first stage. For comparison, we implemented conventional Bayesian reconstruction as described in Eq. (4) using the BV model's prediction as the initial value (denoted as DL-IR). For DLBayesian, we implemented two versions: one finetuned all layers of the prior model (DLBayesian_All) while the other finetuned only the eight most sensitive layers (DLBayesian_Sig8), following Eq. (9) and Eq. (5). The learning rates for DL-IR and DLBayesian were optimized accordingly to ensure stable and efficient convergence. To simplify the comparison, the objective function in the adaptation stage solely focuses on the data fidelity term and drops the regularization term ($\lambda = 0$).

### A. Significance Evaluation

To investigate the correlation of the contribution from layers between the training set and unseen data, we calculated the significance score of each layer in a pretrained DL model for $\Omega_{\text{Train}}$ and $\Omega_{\text{Val}}$ based on the supervised loss in the prior model training stage (see **Fig. 3**). The similar trends observed in the two curves for $\Omega_{\text{train}}$ and $\Omega_{\text{Val}}$ implies that the DL network employs a comparable mechanism to fulfill the same task across different data. Notably, the significance score curve associated with $\Omega_{\text{Train}}$ (see the *y*-axis on the left) exhibits a lower magnitude compared to $\Omega_{\text{Val}}$ (see the *y*-axis on the right). As detailed in Section II.D, the significance score mainly depends on the gradient term and the weights of layers. Since both curves were generated from the same DL model with the same weights, the observed difference in magnitude can be attributed to the gradient associated with different datasets. This observation indicates that the model parameters in stage 1 may be suboptimal for unseen data. Employing the significance score as a metric to guide Bayesian adaptation is a logical solution to precisely tailor model parameters to unseen data.

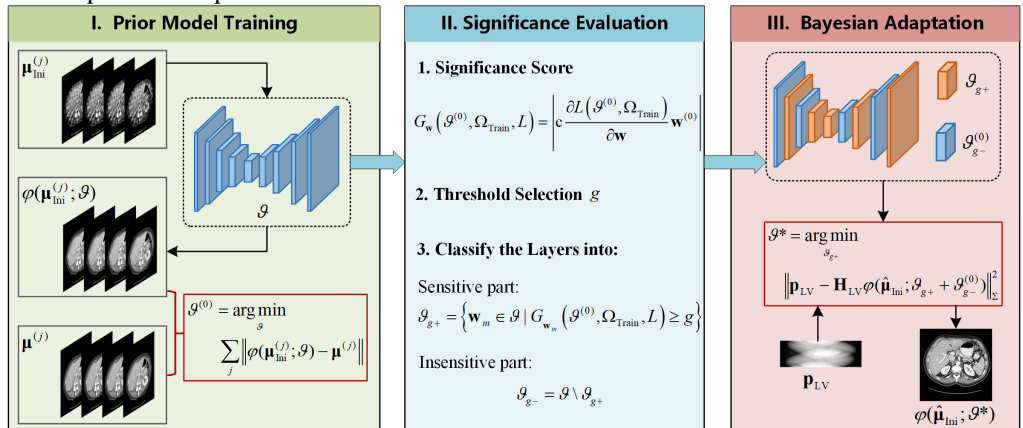

**Fig. 2** Illustration of DLBayesian framework.

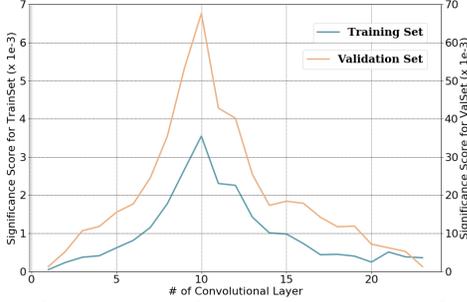

**Fig. 3** Significance scores of layers in the prior model for the training set $G_{\mathbf{w}}(\mathcal{G}^{(0)}, \Omega_{\text{Train}}, L)$ (the blue curve) and the validation set $G_{\mathbf{w}}(\mathcal{G}^{(0)}, \Omega_{\text{Val}}, L)$ (the orange curve).

### B. Experiments on a Circular Sparse-View CT

A circular sparse-view CT scan was simulated with 90 views uniformly distributed over $2\pi$. The distances from the focal spot to the detector and isocenter were 118.5 cm and 78.5 cm, respectively. The detector arrays contain 960 elements equally spaced at 0.45 mm. The reconstruction resolution is $512 \times 512$ with a pixel size of 0.5 mm.

#### 1) Adapt to Different Patients

To evaluate the adaptation performance on different patients, three torso slices are presented in **Fig. 4**. The FBP reconstruction exhibits severe streaking artifacts, whereas FBPConvNet mitigates the artifacts to a large extent with obvious structural distortions. The three adaptation-based methods yielded comparable results with significant improvement in contextual structures and quantitative metrics. However, DL-IR results are blurry in some regions with rich details (see the zoom-ins in **Fig. 4**(D)) while these structures are clearly resolved by the DLBayesian methods (see the zoom-ins in **Fig. 4**(E-F)).

We further demonstrated the convergence behavior of the three adaptation-based methods (see **Fig. 5**) with the iterations of the results in **Fig. 4** marked by the red arrows. Among all the methods, DL-IR reaches the lowest fidelity loss. The performance of DLBayesian is influenced by the constraints imposed by the prior model during data fidelity optimization. Its superior results confirm the effectiveness of the pretraining process before Bayesian adaptation. Further, DLBayesian_Sig8 converges faster than DLBayesian_All, which indicates the necessity of the optimization strategy based on significance scores. A more detailed analysis is in the ablation study.

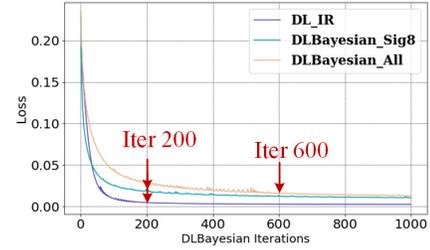

**Fig. 5** Convergence curves in the adaptation process.

#### 2) Adapt to Different Anatomic Structures

We evaluated on head slices to assess the capability of DLBayesian in adapting to different anatomic structures. Notably, the DL networks were trained only on torso data in the first stage. As shown in **Fig. 6**, the direct output of FBPConvNet exhibits severe streaking artifacts due to the distinct anatomic differences. While DL-IR suppresses some residual artifacts left by FBPConvNet, the image quality is obviously inferior compared to the torso slices shown in **Fig. 4**.

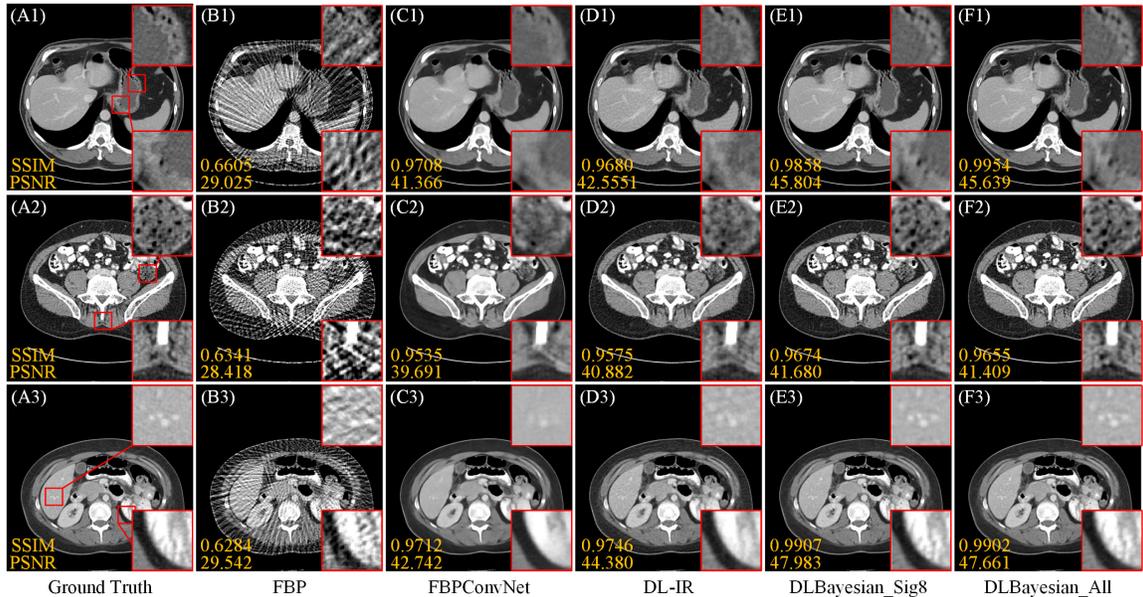

**Fig. 4** Reconstructions of three torso slices: (A) ground truth, (B) FBP, (C) FBPConvNet, (D) DL-IR, (E) DLBayesian_Sig8, and (F) DLBayesian_All. The display window is [0.016, 0.024]. SSIM and PSNR are in the bottom-left corner of each image.

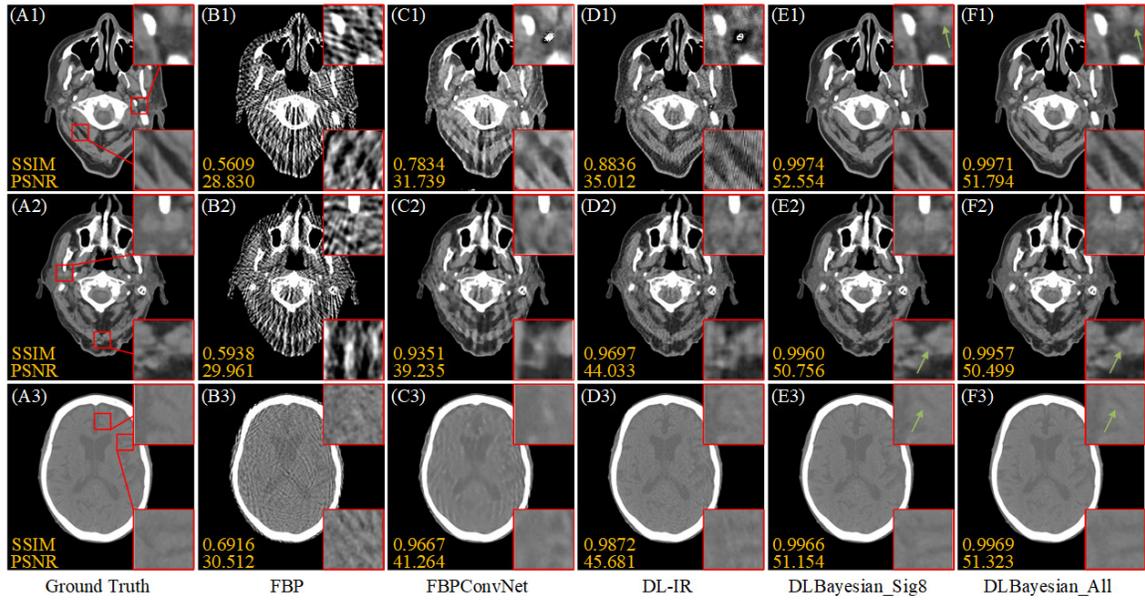

**Fig. 6** Reconstructions of three head slices: (A) ground truth, (B) FBP, (C) FBPConvNet, (D) DL-IR, (E) DLBayesian_Sig8, and (F) DLBayesian_All. The display window is [0.016, 0.024]. SSIM and PSNR are in the bottom-left corner of each image.

This indicates the limitations of DL-IR when facing poor-quality outputs from FBPConvNet. In contrast, DLBayesian remarkably improves the performance in the adaptation stage, as evidenced by the substantial reduction in artifacts and the enhanced clarity of details (see the green arrows in **Fig. 6**(E-F)). This proves the strong capability of DLBayesian in combining data-driven priors and data consistency.

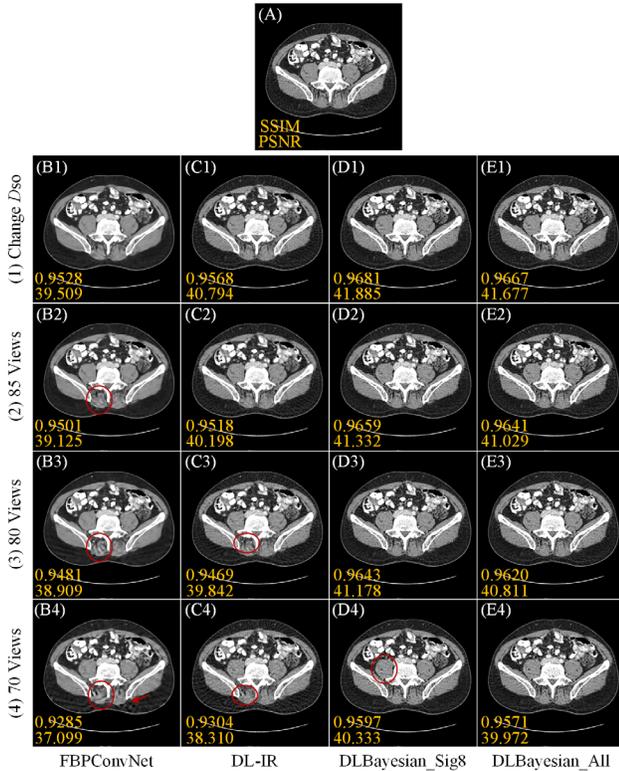

**Fig. 7** Reconstructions of a torso slice under different data sampling conditions: (A) ground truth, (B) FBPConvNet, (C) DL-IR, (D) DLBayesian_Sig8, and (E) DLBayesian_All. There are four scenarios: (1) change the $D_{SO}$ to 60 cm, (2)-(4): reduce the number of projection views to 85 / 80 / 70 views. The display window is [0.016, 0.024]. SSIM and PSNR are in the bottom-left corner of each image.

### 3) Adapt to Different Data Sampling Scenarios

To explore the effectiveness of the methods under different data sampling conditions, we modified the simulated system configuration and tested on the same slice as the second sample in **Fig. 4**. Firstly, we adjusted the distance from the source to the isocenter $D_{SO}$ from 78.5 cm to 60.0 cm and correspondingly extended the number of detector elements from 960 to 1280. All methods produce high-quality images (see **Fig. 7**(1)) which indicates DL models are well-suited to accommodate such changes. Next, we reduced the projection views from 90 to 85, 80, and 70. Fewer views lead to more severe streaking artifacts in conventional FBP reconstructions, and the pretrained FBPConvNet is unable to fully mitigate these artifacts (see **Fig. 7**(B2)-(B4)). For 85 projection views, both DL-IR and DLBayesian successfully restore image quality by leveraging the strength of data fidelity (see **Fig. 7**(C2)-(E2)). However, with the number of projection views further reduced to 80 and 70, DL-IR fails to fully correct the abnormal structures appeared in FBPConvNet results (see the red circles in **Fig. 7**(C3/4)) and produced obvious streaking artifacts while DLBayesian still yields high-quality reconstructions (see **Fig. 7**(D/E3)-(D/E4)). Quantitative metrics also confirm the advantage of DLBayesian in adapting to varying data sampling scenarios.

### C. Experiments on a Multi-segment Linear CT Scan

To further investigate the performance under more general scan trajectories, we tested the methods on a multi-segment linear trajectory using the torso dataset described in Section III.B. As depicted in **Fig. 8**(a), the system consists of a carbon nanotube (CNT) multi-spot X-ray source, a detector, and a turntable. Each source segment consists of 9 focal spots distributed linearly with a spacing of 30 mm. We simulated an 11-segment linear CT scan with 99 projection views unevenly distributed over $2\pi$ (see **Fig. 8**(b)) which introduces a mixed limited-angle and sparse-view problem. The distances from the isocenter to the source and the detector were 68.2 cm and 46.2 cm. The detector consists of 960 elements with a 1 mm pitch.

The reconstruction resolution is 512×512 with a pixel size of 0.6 mm. Previously, we derived an analytical reconstruction algorithm for the multi-segment linear CT under the differentiated back-projection filtration (DBF) framework[7]. For the experiments here, an image-domain U-Net (termed DBFConvNet) is trained to suppress artifacts in the DBF reconstructions. Other experimental settings follow those described in Section III.B.

DBFConvNet output as an initial estimate, DL-IR leverages data consistency to recover contextual structures. However, this process introduces streaking artifacts as indicated by the red arrows in **Fig. 9** (D). The proposed DLBayesian methods efficiently adapt the DBFConvNet to variations in patients and anatomic structures with the best visual quality and quantitative metrics compared to all other methods. These results validate the capability of DLBayesian in addressing the complex data distributions in non-standard trajectories.

### D. Results from a Real Scan of a Dead Rat

To explore the practical performance of DLBayesian, a real scan of a dead rat was performed. The tube voltage and current were 80 kV and 80 mA. The distances from the focal spot to the detector and isocenter were 93.4 cm and 60 cm, respectively. The detector size was 0.15 mm. Projection data were acquired from 750 views over $2\pi$, and subsets of 50/75/125 evenly distributed views were selected to assess reconstruction performance. The FBP reconstruction using all 750 projection views served as the reference image. The reconstruction resolution is 512×512 with a 0.15 mm pixel size. It is worth noting that the system configuration for this practical scan is quite different from the simulation setup in Section III.B. The DL model trained in Section III.B on torso data was directly

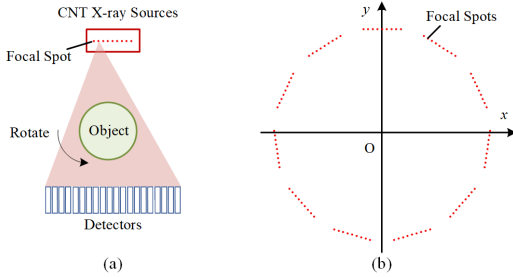

**Fig. 8** Illustration for the multi-segment linear scan trajectory: (a) imaging geometry and (b) distribution of focal spots.

In **Fig. 9**, we display the same torso and head slices shown in **Fig. 4** and **Fig. 6**. DBF reconstructions exhibit severe streaking artifacts due to the non-uniform properties of missing data which significantly differ from the circular trajectory. Although DBFConvNet helps mitigate the artifacts, notable structural distortions remain, particularly in the head slice. By using the applied to reconstruct the rat images (see **Fig. 10**). With 125 projection views, the FBP result (see **Fig. 10**(B1)) has much fewer streaking artifacts compared to the simulation experiment (see **Fig. 4**(B)). These artifacts can be well suppressed by FBPConvNet (see **Fig. 10**(C1)). DL-IR further recovers detailed structures (see the green arrow in **Fig. 10**(D1)) but also introduces new streaking artifacts (indicated by the red circle in **Fig. 10**(D1)). DLBayesian achieves the best image quality with the clearest details (see the green circles in **Fig. 10**(E1)-(F1)). DLBayesian_Sig8 avoids introducing new streaking artifacts as DLBayesian_All. As the number of projection views decreases, FBP results exhibit more severe artifacts (see **Fig. 10**(B2)-(B3)). FBPConvNet mitigates some of these artifacts, but introduces abnormal structures (see the red circles in **Fig. 10**(C2)-(C3)), which significantly degrades the performance of DL-IR (see **Fig. 10**(D2)-(D3)). In contrast, DLBayesian effectively suppresses artifacts by leveraging prior knowledge from the training set though its performance declines as the individual information from the projection data decreases with fewer projection views. These experiments demonstrate the superiority of DLBayesian in managing complex scenarios in practical CT scans.

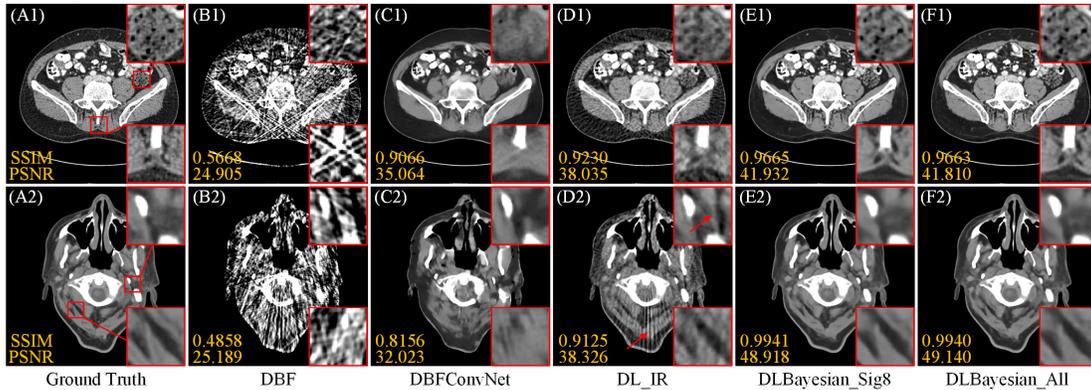

**Fig. 9** Reconstructions of a torso slice and a head slice: (A) ground truth, (B) DBF, (C) DBFConvNet, (D) DL-IR, (E) DLBayesian_Sig8, and (F) DLBayesian_All. The display window is [0.016, 0.024]. SSIM and PSNR are in the bottom-left corner of each image.

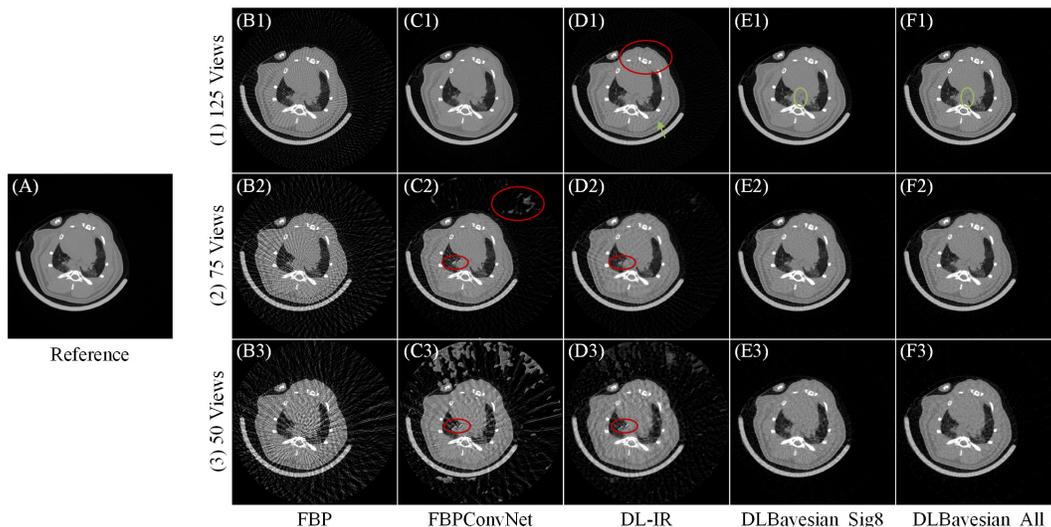

**Fig. 10** Reconstructions for a real scan of a dead rat from different methods: (A) referenced image, (B) FBP, (C) FBPConvNet, (D) DL-IR, (E) DLBayesian_Sig8, and (F) DLBayesian_All. The display window is [0, 0.05].

### E. Ablation Study

We conducted studies on the impact of three key factors in DLBayesian's performance.

#### 1) Impact of the Finetuning-Layer-Selection Strategy

The performance of the DL model is more significantly influenced by its sensitive parts than its insensitive parts. To efficiently adapt the prior model to individual cases, DLBayesian finetunes the eight most sensitive layers. We compared the reconstruction performance of the same slice shown in **Fig. 6** by finetuning the eight most sensitive layers (*Sig8*), and the eight most insensitive layers (*InSig8*) using the same learning rate of $5 \times 10^{-5}$. The reconstructed images and convergence curves are displayed in **Fig. 11** where the iterations of the results presented in **Fig. 11**(a) and (b) are indicated by the red arrows in **Fig. 11**(c). It is evident that *Sig8* achieves better image quality and faster convergence compared to *InSig8*. This validates the importance of finetuning the most sensitive layers for efficient adaptation.

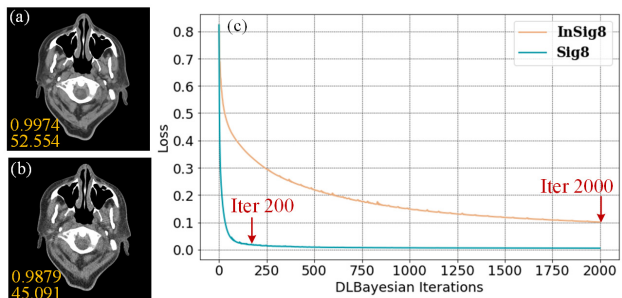

**Fig. 11** Performance on finetuning the sensitive and insensitive parts of the model: reconstructions from Sig8(a), InSig8(b), and the convergence curves(c). The display window for images is [0.016 0.024]. SSIM and PSNR are in the bottom-left corner of each image.

#### 2) Impact of the Number of Finetuned Layers

We evaluated the performance of finetuning 4/8/12/16/23(All) layers of the prior model. As shown in **Fig. 12**, all pipelines achieve strong performance within the DLBayesian framework while finetuning fewer layers leads to faster converges. In the presented case, finetuning eight layers yielded slightly better quantitative metrics. To examine whether the models "forget" the prior knowledge after adaptation, we assessed the performance of the finetuned models on the training set. As shown in TABLE I, finetuning a small portion of the DL model better preserves the performance on the training set. Considering reconstruction quality and computational efficiency, we finetuned 8 layers for our experimental study.

TABLE I
PERFORMANCE OF THE FINETUNED MODELS ON THE TRAINING SET

| # of Tuned Layers | 0 | 4 | 8 | 12 | 16 | 23 |
|---|---|---|---|---|---|---|
| RRMSE in $\Omega_{Train}$ (%) | 1.975 | 2.151 | 2.158 | 2.195 | 2.222 | 2.223 |

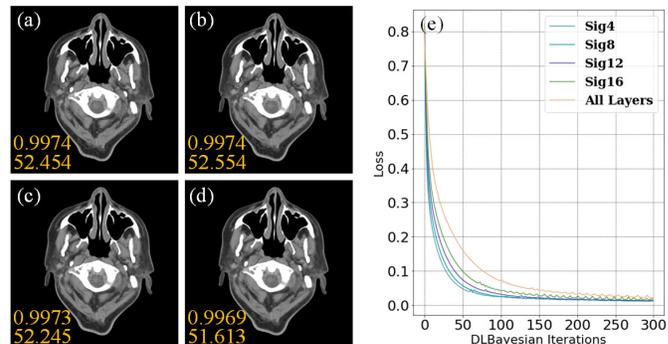

**Fig. 12** Reconstruction performance by finetuning different numbers of layers: (a)-(d) reconstructions by finetuning 4/8/12/16 layers of the prior model, and (e) the convergence curves for different finetuning strategies. The display window for reconstructed images is [0.016 0.024]. SSIM and PSNR are in the bottom-left corner of each image.

#### 3) Impact of the Initialization Strategy

Unlike the existing best-validation mechanism, DLBayesian uses a converged model to maximize learning from the training data. To explore the impact of the initialization strategies, we trained a model on the torso dataset as in Section III.B and evaluated it on the head slice in **Fig. 6** using three initialization strategies: (a) BV_Head and (b) BV_Torso which are the BV models selected based on validation using the head and torso datasets, and (c) the converged model which achieves optimal performance on the training set. Due to the distinct differences between the torso and head slices, the model is seen overfitting at an early stage when the head validation set is used with BV_Head and BV_Torso undergoing training for 4 and 15 epochs, respectively. We first compared the performance of the three models before adaptation. BV_Head achieves better quantitative metrics than BV_Torso (see **Fig. 13**(a-b)) though BV_Torso more efficiently mitigates artifacts (see ROI1). However, BV_Torso reconstruction exhibits abnormal mosaic patterns due to the mismatched data distributions (see ROI2). The converged model achieves the highest quantitative metrics but introduces some fake structures (see the red circle in **Fig. 13**(c)). After Bayesian adaptation, BV_Torso outperforms BV_Head as illustrated in **Fig. 13**(d) and (e). We conjecture that this improvement is due to BV_Torso undergoing more training epochs than BV_Head, which helps to capture more useful information for artifact reduction. This highlights the essential difference between direct generalization and our adaptation approach. Nevertheless, both models fall short in artifact reduction compared to the converged model. The convergence curves in the adaptation stage reveal that the converged model more rapidly reduces the fidelity loss to a lower level (see **Fig. 13**(g)). This confirms the critical importance of employing a converged model as the initial point.

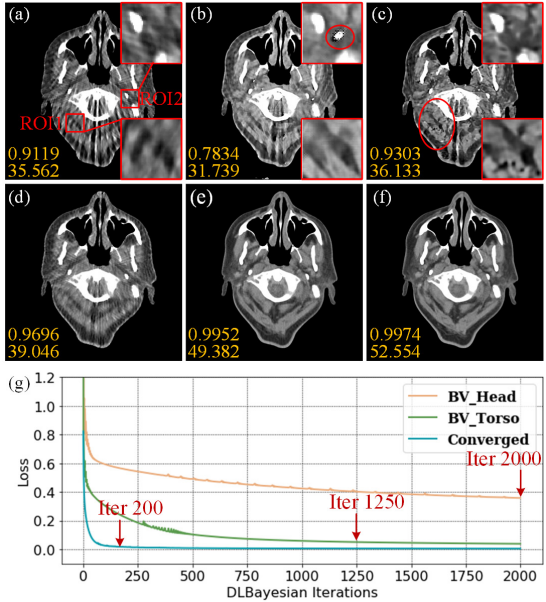

**Fig. 13** Performance of different initial points: (a)-(c) direct inference from the pretrained BV_Head, BV_Torso, and converged models, (d)-(f) DLBayesian reconstructions initialized by BV_Head, BV_Torso, and converged models, and (g) convergence curves for Bayesian adaptation with different initializations. The display window for images is [0.016 0.024]. SSIM and PSNR are in the bottom-left corner of each image.

## IV. DISCUSSION

CT reconstruction with a very limited number of projection views is challenging. DL networks are effective at embedding features and implicitly extracting knowledge while their ability to generalize to unseen data is often a big concern. In this work, we proposed a new method DLBayesian which embraces a "group-to-individual" strategy: it first undergoes supervised learning on an expansive dataset, and then finetunes the network parameters to adapt prior knowledge to individual cases under the guidance of the imaging model and projection data. DLBayesian provides a new perspective in bridging data-driven priors and data consistency and establishes an efficient optimization scheme for practical applications.

Previous methods mainly follow a "Fitting + Generalization" pipeline where the trade-off between fitting the training set and generalizing to unseen data is a big problem. The BV model is a compromise by using the performance on a validation dataset as a critique. However, the composition of the validation dataset can significantly influence the performance of BV models (see **Fig. 13**(a) and (b)). If the real data is hard to access, the validation dataset may not be sufficiently representative which makes the selection of an appropriate BV model difficult. In contrast, DLBayesian adopts a "Fitting + Adaptation" pipeline that separates prior model training and adaptation to the individual case into two independent steps. This separation allows the fitting process to focus entirely on learning from the training data to achieve a converged model for subsequent adaptation. Different from the generalization pipeline, learning from the training data more thoroughly sets a more stable initial point and provides more useful information for the adaptation stage, particularly when the objective tasks and imaged objects share similarities with the training data(see **Fig. 13**(d-f)).

In formulating efficient optimization within the alternative Bayesian reconstruction framework, the optimized network parameters define the solution space of the problem. To delve deeper into the working mechanism of DL networks, we propose to use a significance score as a quantitative metric to assess the contribution from different layers in the model. Our findings reveal that the relative significance relationships among layers remain consistent for both the training set and unseen data (see **Fig. 3**) even though the weights of the network may not be optimal for the unseen data. Hence, the significance score serves as a guide for adaptation where the sensitive part of the prior model is finetuned and the insensitive part is frozen. These frozen layers act as parameter constraints to enhance the adaptation capability in terms of resultant image quality and convergence speed (see **Fig. 4** and **Fig. 5**). Interestingly, after adaptation, the performance on the training set is also better preserved (see TABLE I) by finetuning only the most sensitive layers. This highlights the mechanism of DLBayesian: comprehensively inherit useful prior knowledge and efficiently revise the unsuitable parts according to individual cases.

Our experimental results demonstrate that DLBayesian significantly enhances the adaptability of DL networks across various patients (see **Fig. 4**) and anatomic structures (see **Fig. 6**). Experiments on a special multi-segment linear trajectory and the real scan of a dead rat confirm the flexibility of DLBayesian for a general scan trajectories and the capability to handle complicated variations (protocols, scanned objects, and angular sampling) in practical CT scans (see **Fig. 9** and **Fig. 10**). DLBayesian holds the potential for the reconstruction of new imaging systems and scenarios where obtaining a large-scale training dataset is difficult or even infeasible. However, when the deviation between the training set and real data is relatively large, the performance may degrade with some abnormal structures and secondary artifacts (see the red circle in **Fig. 7**(D4)). In future work, we aim to integrate more DL backbones and enhanced imaging physics models into DLBayesian to better capture prior knowledge and leverage individual information from projection data. This will further broaden the application of DLBayesian to a wider range of system configurations and tasks.

## V. CONCLUSION

We proposed an alternative Bayesian reconstruction framework (DLBayesian) that provides an innovative solution to combining the strengths of DL networks and Bayesian reconstruction. By adopting a "group-to-individual" strategy, DLBayesian effectively adapts the knowledge learned from a large-scale training dataset to individual cases. We designed a new adaptation mechanism guided by a significance score, a quantitative metric proposed to assess the contribution of different layers within the model, which significantly improves the efficiency of adaptation. Our experimental results underscore DLBayesian's superior performance in adapting to diverse patients and anatomic structures. The results also confirmed the flexibility and capability of DLBayesian in handling complex scenarios in practical CT scans.